\documentclass[aps,preprint,groupedaddress,nofootinbib,superscriptaddress,showkeys,showpacs,eqsecnum,amsmath,amssymb]{revtex4-1}
\usepackage{amsfonts}
\usepackage{bm}
\usepackage[colorlinks=true,
    linkcolor=blue,
    citecolor=blue,
    urlcolor=blue]{hyperref}
\usepackage{graphicx}
\usepackage{subcaption}

    \newcommand{\ernst}{\mathcal{E}}
\renewcommand\Re{\operatorname{Re}}
\renewcommand\Im{\operatorname{Im}}
\DeclareMathOperator\arctanh{arctanh}

\begin{document}

\title{Exact rotating wormholes via Ehlers transformations} 

\author{Adolfo Cisterna}
\email{adolfo.cisterna.r@mail.pucv.cl}
\affiliation{Sede Esmeralda, Universidad de Tarapac\'a, Av. Luis Emilio Recabarren 2477, Iquique, Chile}

\author{Keanu Müller}
\email{keanumuller2016@udec.cl}
\affiliation{Departamento de F\'isica, Universidad de Concepci\'on, Casilla, 160-C, Concepci\'on, Chile}

\author{Konstantinos Pallikaris}
\affiliation{Laboratory of Theoretical Physics, Institute of Physics, University of Tartu, W. Ostwaldi 1, 50411 Tartu, Estonia.}
\email{konstantinos.pallikaris@ut.ee}

\author{Adriano Viganò}
\email{adriano.vigano@unimi.it}
\affiliation{Istituto Nazionale di Fisica Nucleare (INFN), Sezione di Milano
Via Celoria 16, I-20133 Milano, Italy}
\affiliation{Università degli Studi di Milano Via Celoria 16, I-20133 Milano, Italy}

\begin{abstract}

In this paper, we construct exact rotating wormholes using the Ehlers solution-generating technique. This is based on the Ernst description of four-dimensional, stationary, and axially symmetric solutions of the Einstein-Maxwell theory. We adopt the static Barcel\'o-Visser wormhole derived from the Einstein-Maxwell-conformal-scalar theory as a seed and demonstrate, through the Ernst approach, how to construct two novel geometries of rotating wormholes. These geometries correspond to the Barcel\'o-Visser wormhole embedded within a rotating and a magnetic background, respectively. In the first case, the rotation is a result of a dragging force (due to the rotating background) acting on the initial static wormhole, while in the second case, it is caused by the electromagnetic interaction between the electric charge of the static wormhole and the external magnetic field. We conduct a comprehensive analysis of the geometric properties of these configurations and examine the new features introduced by rotation, such as the emergence of ergoregions. Recent evidence suggests that incorporating slow rotation can stabilize wormholes, rendering these exact, fully rotating solutions particularly appealing. 
\end{abstract}

\pacs{04.50.Kd, 04.20.Cv, 04.40.Dg}

\maketitle

\section{Introduction}
Soon after the advent of the theory of General Relativity (GR) \cite{Einstein:1915ca,Einstein:1915by}, its first exact solution, the Schwarzschild spacetime \cite{Schwarzschild:1916uq}, was discovered. This spacetime is commonly represented in coordinates adapted to distant static observers, which are not extendable across the event horizon. The complete analytic extension of the Schwarzschild metric was achieved by utilizing null geodesic coordinates, a task that took over 40 years to accomplish \cite{Kruskal:1959vx}. The Schwarzschild geometry comprises four regions: the Schwarzschild black hole, the Schwarzschild white hole, the Schwarzschild patch, and its mirror spacetime. By utilizing isotropic coordinates, it was discovered that the last two regions are connected through the Einstein-Rosen bridge \cite{Einstein:1935tc}, which, although not traversable, laid the foundation for the conjecture and exploration of the theoretical and experimental existence of wormholes. Wormhole spacetimes are visualized as tunnels connecting two different universes or two distant regions of the same universe through a throat, which necessitates the presence of matter with exotic energy behavior being traversable\footnote{For a wormhole to be traversable, the throat must be robust enough to allow the transfer of matter without collapsing. This requires the presence of matter around the throat that violates standard energy conditions. For a comprehensive introduction to the subject, we refer readers to Ref. \cite{Visser:1995cc}.}. The study of wormholes has greatly benefited from the initial guidelines provided by Morris and Thorne \cite{Morris:1988cz}. They have characterized the geometric features that a given spherically symmetric spacetime should exhibit to represent a wormhole and the energy conditions that the corresponding source should meet for the throat to be traversable. In recent decades, the study of wormholes has received substantial attention, not only at a theoretical level but also in the field of astrophysics. The analysis of their shadows \cite{Nedkova:2013msa,Ohgami:2015nra,Shaikh:2018kfv}, lensing effects \cite{Harko:2008vy,Lamy:2018zvj}, or possible accretion disks around them \cite{Harko:2009xf,Zhou:2016koy} could provide experimental evidence of their existence in nature. Furthermore, in light of the recent detection of gravitational waves \cite{LIGOScientific:2016aoc,LIGOScientific:2017vwq}, the study of wormhole quasinormal modes \cite{Konoplya:2018ala,Blazquez-Salcedo:2018ipc}, which is less explored than the black hole case, is particularly appealing.\\

Wormhole geometries have been discovered in a wide range of models, both in GR with exotic matter of various types \cite{Hochberg:1996ee,Konoplya:2021hsm,Lobo:2005us,Bronnikov:2005gm}, and in modified theories of gravity where additional degrees of freedom provide the necessary setup \cite{Lobo:2009ip,Antoniou:2019awm,Bronnikov:2013coa,Bronnikov:2015pha,Korolev:2014hwa,Harko:2013yb,Rahman:2023swz}. In addition, recent explorations considered the construction of charged higher-dimensional wormholes by properly connecting two copies of anti-de Sitter, where the charged is supported by higher-dimensional $p$-forms \cite{Deshpande:2022zfm}.
However, the majority of these solutions considers static spacetimes. The examination of exact rotating solutions is significantly more complex, and in the context of black holes and wormholes, only a limited number of geometries other than the Kerr black hole are known \cite{Stephani:2003tm}. In a previous study \cite{Teo:1998dp}, Teo provided a rotating extension of the Morris and Thorne work, outlining the conditions that a general stationary and axially symmetric spacetime must meet in order to represent a rotating wormhole. It was demonstrated that the presence of exotic matter remains an essential component, even if it can be localized in a specific sector of the would-be throat, and that ergoregions take place. However, these ergoregions are located around the wormhole throat, and their shape is significantly different from that in the black hole case. No event horizons are present. Only a few novel rotating wormhole backreactions are known, both numerically in the case of fast rotation and perturbatively in the slow rotation regime \cite{Kashargin:2007mm,Kashargin:2008pk,Kleihaus:2014dla,Chew:2016epf}\footnote{Here, we mean that only a few new backreactions are known in this context \cite{Matos:2010pcd,DelAguila:2015isj}; however, known spacetimes that acquire the interpretation of rotating wormholes have been previously analyzed \cite{Clement:2022pjr,Anabalon:2018rzq}.}.

Beyond its intrinsic significance, rotation may also play a role in the stability of this type of configurations. It has been proposed in the past that rotation could stabilize static wormholes \cite{Matos:2005uh}, and recently, Azad \emph{et al}.~\cite{Azad:2023iju} have provided direct evidence that rotation, even at a perturbative level, might induce the dynamical stabilization of four-dimensional wormholes. The authors have shown that the known radially unstable Ellis-Bronnikov wormhole \cite{Ellis:1973yv,Bronnikov:1973fh} might become stable when generalized to its second-order perturbative rotating extension. This makes the study of rotating wormhole geometries particularly attractive.\\

In this paper, we propose a novel methodology for constructing exact rotating wormholes as a means of evaluating their dynamical stability, in line with the recent results of Ref. \cite{Azad:2023iju}. 
In contrast to the Teo ansatz \cite{Teo:1998dp}, where rotation is given by the angular momentum of a rotating source, we propose introducing rotation to a static wormhole through two distinct mechanisms. First, we consider embedding a static uncharged wormhole geometry within a rotating background spacetime, known as swirling geometry \cite{Astorino:2022aam}, which possesses characteristics similar to those of a Kerr-like rotating metric. Consequently, the embedded geometry rotates via a dragging effect due to the rotating background. On the other hand, we propose embedding a charged static wormhole in a Melvin Universe \cite{Ernst:1976mzr}, in which the rotation is acquired through the interaction of the wormhole's charge with the external magnetic field. To carry out these embeddings, we will utilize the Ernst scheme for electrovacuum \cite{Ernst:1967wx,Ernst:1967by}, which allows us to identify and use a set of Lie point symmetries of the Einstein-Maxwell system known as Ehlers transformations \cite{Ehlers:1957zz,Harrison:1968}. These symmetries can convert a given stationary axially symmetric solution of the Einstein-Maxwell equations into a new, nonequivalent solution, providing us with a powerful solution-generating technique to explore new solutions of the Einstein-Maxwell field equations. We will also elaborate on how the original Ernst setup can be extended to Einstein-Maxwell-scalar theories \cite{Astorino:2013xc,Astorino:2014mda} and how this expands the possibility of considering wormhole geometries as seeds. Specifically, we will take a known traversable wormhole, the Barcel\'o-Visser spacetime \cite{Barcelo:1999hq,Barcelo:2000zf}, a solution of a conformally coupled scalar theory, as a seed and use the Ehlers transformations to generate exact rotating swirling and Melvin wormholes.\\
\\
The manuscript is structured as follows. In Sec. II, we go through some preliminary results, necessary for constructing the Barcel\'o-Visser wormhole, and we briefly review the Ernst approach to Einstein-Maxwell theory and Ehlers symmetries. The Barcel\'o-Visser wormhole is then constructed in detail using a solution of the Einstein-scalar system. Connections to the  Fisher-Janis-Newman-Winicour (FJNW) spacetime \cite{Janis:1969ivo} and the Bronnikov-Melnikov-Bocharova-Bekenstein  (BBMB) black hole \cite{BBMB,Bekenstein:1974sf} are highlighted to aid in understanding the solution's geometric properties. 
Section III constructs rotating metrics using Ehlers and Harrison magnetic maps applied to the Barcel\'o-Visser line element, resulting in swirling and magnetic rotating wormholes. 
The solutions' geometric properties, including the position and traversability of the throat and ergospheres, and the asymptotic behavior of the metric and matter fields, are analyzed in detail. Finally, in Sec. IV, these rotating spacetimes are intuitively discussed, and further applications in astrophysical settings are proposed. 

\section{Preliminaries}

Before proceeding with the construction of our solutions, it is necessary to address certain preliminary considerations. One major challenge in our approach is finding a suitable static wormhole spacetime on which to apply the Ehlers transformations \cite{Ehlers:1957zz,Harrison:1968}. The Ehlers transformations, which are Lie point symmetries of the Einstein-Maxwell system, do not hold for solutions of the Einstein-Maxwell-conformal-scalar theory, such as the Barcel\'o-Visser wormhole \cite{Barcelo:1999hq,Barcelo:2000zf}. However, it has been proven in Ref. \cite{Astorino:2013xc} that the Einstein-Maxwell-scalar theory preserves the Ehlers symmetries, thus allowing us to circumvent this obstacle by using Bekenstein transformations \cite{Bekenstein:1974sf}. These transformations involve a conformal rescaling of the metric and a redefinition of the scalar field, enabling us to transform the Barcel\'o-Visser wormhole solution of the Einstein-Maxwell-conformal-scalar theory into a solution of the Einstein-Maxwell-scalar system. With the desired solution in the so-called Einstein frame, the validity of the Ehlers transformation is guaranteed. The last step is to apply an inverse Bekenstein transformation to return the geometry to the conformal frame and, in this manner, obtain the Ehlers-transformed version of the desired solution. This step is crucial, as the causal structure of the spacetime is sensitive to changes of frame. Therefore, the Barcel\'o-Visser wormhole does not represent a wormhole in the Einstein frame; instead, it represents a singular spacetime, a relative of the well-known FJNW solution \cite{Janis:1969ivo}. In the following section, we will review the construction of the Barcel\'o-Visser wormhole starting from the FJNW solution, the most basic solution of the Einstein-scalar system. This spacetime is known to be connected to the BBMB black hole \cite{BBMB,Bekenstein:1974sf} through Bekenstein transformations. The BBMB black hole represents the black hole cousin of the Barcel\'o-Visser wormhole, both being related by a specific metric rescaling and a suitable shift of the scalar field \cite{Ayon-Beato:2015ada}. With these ingredients at hand, we will proceed to a concise review of the Ernst approach of Einstein-Maxwell theory and the corresponding Ehlers transformations. These preliminary steps will enable us to construct rotating swirling and Melvin wormholes.

\subsection{Seed: Constructing Barcel\'o's wormhole}

We start by considering the Einstein-scalar system with action 
\begin{equation}
I[g_{\mu\nu},\phi]=\int d^{D}x\sqrt{-g}\left[  \frac{R}{2\kappa}-\frac{1}%
{2}g^{\mu\nu} \partial_\mu\phi\partial_\nu\phi\right] ,
\label{FJNWaction}%
\end{equation}
where $\kappa\equiv 8\pi G_{D}$ with $G_{D}$ being Newton's gravitational constant in $D$ spacetime dimensions.\footnote{We work with natural units $\hbar=c=1$.} It has been proven that the corresponding field equations, namely,
\begin{subequations}
\begin{align}
G_{\mu\nu}-\kappa\left[\partial_\mu\phi\partial_\nu\phi-\frac{1}{2}g_{\mu\nu}(\partial\phi)^2\right]&=0,\\
\Box\phi&=0,
\end{align}
\end{subequations}
when evaluated for any static vacuum metric of the form 
\begin{equation}
ds^2=-F(r)^2dt^2+F(r)^{-2}h_{ij}dx^idx^j,
\end{equation}
admit the following scalar-enhanced solution,
\begin{subequations}
\begin{align}
ds^2&=-F(r)^{2\alpha}dt^2+F(r)^{-2\alpha}h_{ij}dx^idx^j,\\
\phi&=\sqrt{2}\zeta\alpha\ln F(r), 
\end{align}
\end{subequations}
where $\zeta$ is a dimensionful integration constant with mass dimension $(D-2)/2$, and the dimensionless constant $\alpha$ is related to the Einstein constant $\kappa$ via $\alpha=(1+\kappa\zeta^2)^{-1/2}$. Considering the static vacuum solution to be given by the Schwarzschild metric, it follows that the Einstein-scalar spacetime, in four dimensions, is given by the FJNW solution,\footnote{Please note that we use the symbol $\phi$ to denote the scalar-field profile in the FJNW solution and that the azimuthal coordinate is given by $\vartheta$. We define the transverse codimension-2 manifold to be $d\Omega^2=d\theta^2+\sin\theta^2 d\vartheta^2$ in four dimensions.} which in a suitable gauge reads 
\begin{subequations}\label{janis}
\begin{align} 
     ds^2&=-\left(1-\frac{2m}{r}\right)^A dt^2 + \frac{d r^2}{\left(1-\frac{2m}{r}\right)^A} + \frac{r^2d\Omega^2}{\left(1-\frac{2m}{r}\right)^{A-1} }  \\
    \phi &= \sqrt{\frac{1-A^2}{2 \kappa}} \ln \left(1-\frac{2m}{r}\right).
\end{align}
\end{subequations}
The real parameter $A$ assumes values in $[0,1]$. For $A\neq1$, the solution features a naked singularity on the would-be horizon $r=2m$, whereas it clearly represents the Schwarzschild black hole for $A=1$. 

As shown by Bekenstein \cite{Bekenstein:1974sf}, the Einstein-scalar theory (\ref{FJNWaction}) is conformally related to the Einstein-conformal-scalar theory
\begin{equation}
\bar{I}[\bar{g}_{\mu\nu},\psi]=\int d^Dx \sqrt{-\bar{g}}\left[\frac{\bar{R}}{2\kappa}-\frac{1}{2}\bar{(\partial \psi)}^2 -\frac{\xi_D}{2}\bar{R}\psi^2\right], \label{BBMBaction}
\end{equation}
by means of
\begin{subequations}
\begin{eqnarray}
\bar{g}_{\mu\nu}&=&\cosh^{4/(D-2)}\left(\phi\sqrt{\kappa \xi_D}\right)g_{\mu\nu}, \\
\psi&=&\frac{1}{\sqrt{\kappa\xi_D}}\tanh\left(\phi\sqrt{\kappa\xi_D}\right),
\end{eqnarray}
\label{bektrans}
\end{subequations} 
where $\bar{(\partial \psi)}^2 \equiv \bar{g}^{\mu\nu}\partial_\mu\psi \partial_\nu\psi$ and $\xi_D = ({D-2})/{[4(D-1)]}$. These transformations correspond to a local Weyl rescaling of the metric together with a suitable scalar-field redefinition.\footnote{These transformations are still valid in the presence of a cosmological constant as long as a conformal self-interacting potential for the scalar field is included.} 

The new solution $(\bar{g}_{\mu\nu},\psi)$ should solve the transformed field equations, which in four dimensions read
\begin{subequations}
\begin{align}
0&=\bar{G}_{\mu\nu}-\kappa\left[\partial_\mu\psi\partial_\nu\psi-\frac{1}{2}\bar{g}_{\mu\nu}\bar{(\partial \psi)}^2+\frac{1}{6}\left(\bar g_{\mu\nu}\bar{\Box}-\bar{\nabla}_\mu\bar{\nabla}_\nu+\bar{G}_{\mu\nu}\right)\psi^2\right],\\
0&=\bar{\Box}\psi-\frac{1}{6}\psi\bar{R}.
\end{align}
\end{subequations}
A particular solution of these field equations, the BBMB black hole, is easily obtained by mapping the four-dimensional FJNW solution~\eqref{janis} with $A=1/2$ to the conformal frame. Renaming $m=2M$ and using coordinates $(t,\tilde{r},\theta,\vartheta)$ with
\begin{equation}
    r(\Tilde{r})= \frac{\tilde{r}^2}{\tilde{r}-M},
\end{equation}
it reads  
\begin{subequations}
\label{eq:BBMBbh}
\begin{align}
d\bar{s}^2&=-\left(1-\frac{M}{\bar r}\right)^2dt^2+\frac{d\tilde{r}^2}{\left(1-\frac{M}{\tilde{r}}\right)^2}+\tilde{r}^2d\Omega^2\\
\psi&=\sqrt{\frac{6}{\kappa}}\frac{M}{\tilde{r}-M}.
\end{align}
\end{subequations}
It is then evident that the Bekenstein map severely alters the causal structure of a given spacetime, highlighting the utilization of frame changes as a basic solution-generating technique. 
Now, a very appealing change of frame \cite{Ayon-Beato:2015ada} is the one connecting two Einstein-conformal-scalar theories through the map
\begin{subequations}
\label{MOKtrans}
\begin{align}
\hat{g}_{\mu\nu}&=\left(a \psi \sqrt{\kappa\xi_{D}}+1\right)^{\frac{4}{D-2}}\bar{g}_{\mu\nu},\\
\varphi&=\frac{1}{\sqrt{\kappa\xi_{D}}}\frac{\psi\sqrt{\kappa\xi_{D}}+a}{a\psi\sqrt{\kappa\xi_{D}}+1}, 
\end{align}
\end{subequations}
which drastically enriches the causal-structure spectrum in the new theory.\footnote{This map effectively connects two Einstein-conformal-scalar theories when the cosmological constant and the initial scalar self-interaction are both absent. If both are present, the new theory differs from the original one in the following way: it acquires an effective cosmological constant, and the conformal potential is generalized to include all power-counting super-renormalizable contributions in certain dimensions.}  Here, $a$ is a dimensionless constant related to a shift of the original scalar field $\psi$ with unitarity constraints forcing $a^2<1$ (see Ref.~\cite{Ayon-Beato:2015ada} for more details). Under the action of (\ref{MOKtrans}), the BBMB black hole~\eqref{eq:BBMBbh} transforms into the so-called traversable Barcel\'o-Visser (BV) wormhole~\cite{Barcelo:1999hq,Barcelo:2000zf}
\begin{subequations}
\label{static-worm}
\begin{align}
d\hat{s}^2 &= \left( \frac{\tilde{r}-(1-a)M}{\tilde{r}-M} \right)^2 d\bar{s}^2,\\ 
\varphi &= \sqrt{\frac{6}{\kappa}} \, \frac{a\tilde{r} + (1-a)M}{\tilde{r}-(1-a)M}, 
\end{align} 
\end{subequations}
known for connecting two asymptotically flat regions, $\tilde r=M$ and $\tilde r=\infty$, through a nontrivial throat located at $\tilde{r}=(1+\sqrt{a})M$. Clearly, we further need to consider $0\leq a<1$ in order for the conformal factor to be nonzero in the physical radial domain. This geometry, which we shall discuss in more detail in Sec.~\ref{sec:RotWH}, is the starting point for the construction of our rotating swirling and Melvin wormhole configurations. 

\subsection{Ehlers symmetries: Ehlers and Harrison maps\label{sec:EhlersSym}}

The Ernst equations~\cite{Ernst:1967wx,Ernst:1967by} are a sophisticated formulation of the Einstein-Maxwell equations for stationary and axisymmetric spacetimes. Their structure brings certain interesting symmetries of the Einstein-Maxwell system to the forefront, thereby facilitating the generation of nontrivial solutions. Specifically, the Einstein-Maxwell equations (for stationary and axisymmetric spacetimes) can be equivalently represented by the Ernst equations\footnote{In this section and for the rest of the manuscript, we follow the original convention for $\kappa$ utilized in Refs. \cite{Ernst:1967wx,Ernst:1967by}, namely, $\kappa=2$.}
\begin{subequations}
\label{ernst-eqs}
\begin{align}
\left( \Re\ernst + \big|\Phi\big|^2 \right) \nabla^2\ernst & =
\vec{\nabla}\ernst\cdot \bigl( \vec\nabla\ernst + 2\Phi^* \vec\nabla\Phi \bigr) \,, \\
\left( \Re\ernst + \big|\Phi\big|^2 \right) \nabla^2\Phi & = \vec\nabla\Phi\cdot \bigl( \vec\nabla\ernst + 2\Phi^* \vec\nabla\Phi \bigr) \,,
\end{align}
\end{subequations}
where $\vec{\nabla}$ is the flat differential operator in cylindrical coordinates with gradient and Laplacian given by
\begin{subequations}
    \begin{eqnarray}
        \vec{\nabla} k &=& \vec{e}_\rho \partial_\rho k + \vec{e}_\vartheta\frac{1}{\rho}\partial_\vartheta k + \vec{e}_z \partial_zk,\\
        \nabla^2k &=& \frac{1}{\rho}\partial_\rho (\rho \partial_\rho k) + \frac{1}{\rho^2}\partial_\vartheta \partial_\vartheta k + \partial_z\partial_z k,
    \end{eqnarray}
\end{subequations}
respectively, for any scalar function $k(\rho,z,\vartheta)$. Here, $\partial_{x^i}\equiv \partial/\partial x^i$, and $\{\vec{e}_{x^i}\}_{i=1,2,3}$ is the standard basis of $\mathbb{R}^3$.

The Ernst potentials, defined as
\begin{equation}
\ernst = f - |\Phi|^2 + i \chi , \qquad
\Phi = A_t + i\tilde{A}_\vartheta ,
\end{equation}
are complex scalar functions constructed from the characteristic functions of a general stationary and axisymmetric spacetime, as represented by the Lewis-Weyl-Papapetrou (LWP) metric, and of a stationary and axially symmetric Maxwell field, namely,
\begin{subequations}
\label{lwp-e}
\begin{align}
{ds}^2 & =  f ( dt - \omega d\vartheta )^2 - f^{-1} \bigl[ \rho^2 {d\vartheta}^2 + e^{2\gamma}  \bigl( {d \rho}^2 + {d z}^2 \bigr) \bigr] \,, \\
A & = A_t dt + A_\vartheta d\vartheta \,,
\end{align}
\end{subequations}
respectively, where all functions depend on $(\rho,z)$. The twisted potentials $\tilde{A}_\vartheta$ and $\chi$ are defined via the differential equations 
\begin{equation}
\vec{e}_\vartheta \times \vec\nabla \tilde{A}_\vartheta =
\rho^{-1} f \left( \vec{\nabla} A_\vartheta + \omega \vec{\nabla} A_t \right) \,,
\end{equation}
and
\begin{equation}
\label{chi}
\vec{e}_{\vartheta} \times \vec{\nabla}\chi =
-\rho^{-1} f^2 \vec{\nabla}\omega -2 \vec{e}_{\vartheta} \times \Im\left( \Phi^* \vec\nabla\Phi \right) \,.
\end{equation}
Stationary and axially symmetric spacetimes possess two commuting Killing vectors, ($\partial_t$, $\partial_\vartheta$), and three arbitrary functions depending on the non-Killing coordinates $(\rho,z)$. It is important to mention that, due to the integrability features of the system, the function $\gamma(\rho,z)$ decouples from the others, and it can be uniquely determined by $f(\rho,z)$ and $\omega(\rho,z)$. 

The LWP metric~\eqref{lwp-e} does not represent the unique ansatz for a general stationary and axisymmetric spacetime. In fact, there is another nonequivalent ansatz we can consider, related to the previous one through a discrete double Wick rotation, $t \to i \hat{\vartheta}$ and $\vartheta \to i\hat{t}$. Renaming $\hat{t}=t$ and $\hat{\vartheta}=\theta$, it reads 
\begin{subequations}
\label{lwp-m}
\begin{align}
{ds}^2 & =
-f (d\vartheta - \omega dt)^2
- f^{-1} \bigl[ e^{2\gamma} \bigl( {d\rho}^2 + {dz}^2 \bigr)
- \rho^2 {dt}^2 \bigr] \,, \\
A & = A_t dt + A_\vartheta d\vartheta \,,
\end{align}
\end{subequations}
where now $\Phi = A_\vartheta+ i\tilde{A}_t$ and
\begin{equation}
\label{At}
\vec{e}_{\vartheta} \times \vec\nabla \tilde{A}_t =
\rho^{-1} f \bigl( \vec\nabla A_t + \omega \vec\nabla A_\vartheta \bigr) \,.
\end{equation}
The ansatz~\eqref{lwp-m} gives rise to the very same Ernst equations~\eqref{ernst-eqs}. The main feature of the 
Ernst equations is that they enjoy a set of Lie point symmetries, known as Ehlers symmetries~\cite{Stephani:2003tm},
\begin{subequations}
\label{ernst-group}
\begin{align}
\label{gauge1}
\ernst & = |\lambda|^2 \ernst_0 \,, \qquad\qquad\qquad\quad\;\;
\Phi = \lambda \Phi_0 \,, \\
\label{gauge2}
\ernst & = \ernst_0 + ib \,, \qquad\qquad\qquad\quad\,
\Phi = \Phi_0 \,, \\
\label{ehlers}
\ernst & = \frac{\ernst_0}{1 + i\jmath\ernst_0} \,, \qquad\qquad\qquad\;\,
\Phi = \frac{\Phi_0}{1 + i\jmath\ernst_0} \,, \\
\label{gauge3}
\ernst & = \ernst_0 - 2\beta^*\Phi_0 - |\beta|^2 \,, \qquad\quad\!
\Phi = \Phi_0 + \beta \,, \\
\label{harrison}
\ernst & = \frac{\ernst_0}{1 - 2\alpha^*\Phi_0 - |\alpha|^2\ernst_0} \,, \qquad
\Phi = \frac{\alpha\ernst_0 + \Phi_0}{1 - 2\alpha^*\Phi_0 - |\alpha|^2\ernst_0} \,,
\end{align}
\end{subequations}
where $\alpha$, $\beta$, and $\lambda$ are complex parameters, while $b$ and $\jmath$ are real.
Notice that the only nontrivial transformations are~\eqref{ehlers} and~\eqref{harrison}, the so-called Ehlers~\cite{Ehlers:1957zz} and Harrison~\cite{Harrison:1968} transformations, respectively. Being Lie point symmetries, these transformations leave the Ernst equations unchanged, while simultaneously generating nonequivalent spacetimes. The specific effect of the symmetries depends on which ansatz they are applied to~\cite{Vigano:2022hrg}, i.e., Eq.~\eqref{lwp-e} or~\eqref{lwp-m}. The Ehlers map is known to add a NUT parameter to~\eqref{lwp-e}, whereas it embeds~\eqref{lwp-m} in a rotating (swirling) background~\cite{Astorino:2022aam}. On the other hand, the Harrison map adds a dyonic charge to~\eqref{lwp-e}, while it embeds the seed spacetime~\eqref{lwp-m} in an electromagnetic universe~\cite{Ernst:1976mzr}.

In Ref. \cite{Astorino:2013xc}, it was shown that this set of symmetries still survives after the inclusion of a minimally coupled scalar field, actually, even in the presence of a collection of scalar fields in the context of a sigma model. Denoting the new field by $\Psi$, it is possible to show that all Ehlers symmetries are symmetries of the reduced action principle 
\begin{align}
I[\ernst,\Phi,\Psi] &= \int \rho d\rho dz d\vartheta \Biggl(\frac{(\vec{\nabla}\ernst + 2 \Phi^{*} \vec{\nabla} \Phi ) \cdot (\vec{\nabla}\ernst^{*} + 2 \Phi \vec{\nabla} \Phi^{*} )}{(\ernst+ \ernst^{*} + 2 |\Phi|^2)^2}-\frac{2 \vec{\nabla}\Phi \cdot \vec{\nabla}\Phi^{*}}{\ernst + \ernst^{*} + 2 |\Phi|^2} -\frac{\kappa}{2} \vec{\nabla}\Psi \cdot \vec{\nabla} \Psi \Biggr) , \label{maxw-GR+MinScl-action}
\end{align}
where now the set of symmetries (\ref{ernst-group}) needs to be complemented with the trivial transformation $\Psi_0\rightarrow\Psi=\Psi_0$, while the Ernst equations are complemented with the corresponding Klein-Gordon equation for $\Psi$, $\nabla^2\Psi=0$. 
The Ehlers solution-generating technique, enhanced by the presence of a minimally coupled scalar field, allows for the application of Bekenstein transformations to obtain Ehlers-transformed spacetimes for theories with conformally coupled scalar fields. In the following section, the Ehlers and Harrison maps will be used to generate two rotating versions of the BV wormhole. The Ehlers map will result in the embedding of the BV wormhole into a swirling background, while the Harrison map will be used to embed the electrically charged version of this wormhole into an external magnetic field, resulting in a rotating configuration due to the electromagnetic interaction between the electric monopole and the external magnetic field.

\section{Rotating wormholes\label{sec:RotWH}}

Let us start by considering the BV wormhole \cite{Barcelo:1999hq,Barcelo:2000zf}, which in the presence of electric charge reads 
\begin{subequations}
\begin{align}
d\hat{s}^2 &= \left( \frac{a\sqrt{M^2-\frac{Q^2}{1-a^2}} + r-M}{r-M} \right)^2 d\bar{s}^2,\\
A &= \frac{Q}{r} \, dt, \\
\varphi &= \sqrt{3} \, \frac{\sqrt{M^2-\frac{Q^2}{1-a^2}} + a(r-M)}{a\sqrt{M^2-\frac{Q^2}{1-a^2}} + r-M} \,,
\end{align}
\end{subequations}
with $Q$ representing the monopole electric charge and where $d\bar{s}^2$ was introduced in ~\eqref{eq:BBMBbh}.\footnote{We have renamed $\tilde{r}\equiv r$.} This can be found in Ref.~\cite{Barrientos:2022avi} for zero NUT parameter and cosmological constant, but it can also be directly obtained by acting with transformations (\ref{MOKtrans}) on the \emph{electrically charged} BBMB black hole~\cite{Bekenstein:1974sf}. The key change in the causal structure of the solution is the appearance of a double pole in the conformal factor. This pole defines the position of conformal infinity at $r=M$, thus introducing a second asymptotic region in the spacetime. Therefore, the physical spacetime is the one belonging to the interval $]M,\infty[$ of the radial coordinate, while $t$, $\theta$, and $\vartheta$ maintain their usual interpretation. The new spacetime is then devoid of an event horizon. On the other hand, the standard curvature singularity at $r=0$ is accompanied by another curvature singularity, this time at the locus of points where the conformal factor vanishes. Both curvature singularities exist at $r$ values outside the physical domain of the radial coordinate, and hence they do not pose a threat.

The surface area of the transverse 2-manifold, proportional to the square root of the determinant of the induced 2-metric ($t=\mathrm{const.}=r$), has a minimum at
\begin{equation}
r_{\circ}=(1+\sigma)M,\label{eq:BVthroat}
\end{equation}
where 
\begin{equation}
    \sigma = \sqrt{a}\left(1 - \frac{Q^2}{(1-a^2)M^2}\right)^{1/4}.
\end{equation}
It is interpreted as the radius of a wormhole throat that connects two asymptotically flat regions, one at $r=M$ and another at $r=\infty$. Both scalar fields and gauge fields are wellbehaved in the vicinity of the asymptotic regions. We refer to  Refs. \cite{Barrientos:2022avi,Anabalon:2012tu} for further details about this type of geometries. Moreover, the wormhole is traversable. A radial null observer at the equator and at a fixed arbitrary azimuthal angle traverses the throat in a finite amount of time equal to
\begin{equation}
\delta t= \int_{r_\circ-\epsilon}^{r_\circ + \epsilon}dr \frac{r^2}{(r-M)^2} = 2\left( \epsilon + \frac{\epsilon M^2}{\sigma^2 M^2 - \epsilon^2}+M\ln \frac{\sigma M+\epsilon}{\sigma M -\epsilon}\right),\label{eq:travBVstatic}
\end{equation}
where $\sigma M > \epsilon >0$. Finally, a qualitative study of the energy-momentum components reveals violation of the weak energy condition as one would customarily expect. In what follows, we take advantage of the properties of the BV wormhole, and by making use of the Ehlers symmetries as a solution-generating technique, we construct two families of exact rotating wormholes. 

\subsection{Swirling wormholes}

We start by considering the uncharged BV spacetime~\eqref{static-worm} as a seed on which we will apply Ehlers transformations (\ref{ehlers}). As stated before, when acting on a doubly Wick-rotated axially symmetric configuration, the Ehlers transformations embed the given seed spacetime into a swirling background. However, in order for the transformations to be applicable in the first place, we need to restrict ourselves to Einstein-scalar theory (\ref{FJNWaction}); this is why we first act on the BV configuration~\eqref{static-worm} with the inverses of the Bekenstein transformations (\ref{bektrans}), obtaining (after renaming $\tilde{r}\equiv r$)
\begin{subequations}
\begin{align}
ds^2 &=\left[1 - \left(\frac{M + a(r-M)}{r-(1-a)M}\right)^2\right] \Omega\,
d\bar{s}^2, \\ 
\phi &= \sqrt{3} \, \arctanh\left(\frac{M + a(r-M)}{r-(1-a)M}\right) \,,
\end{align}\label{BVEF} 
\end{subequations}
where 
\begin{equation}
    \Omega(r)=\left( \frac{r-(1-a)M}{r-M}\right)^2.\label{eq:Omega}
\end{equation}

Now, having the Einstein-frame metric, we can proceed with the identification of the seed-metric functions with those in the magnetic LWP ansatz~(\ref{lwp-m}), once we first write the latter in coordinates $(t,r,\theta,\vartheta)$ with 
\begin{equation}
    \rho = \frac{(1-a^2)(r-2M)r}{r-M}\sin\theta,\quad z=\frac{(1-a^2)(r^2-2M r + 2M^2)}{r-M}\cos\theta.
\end{equation}
We thus find that
\begin{subequations}\label{seedmetricLWP}
\begin{align}
f_0(r,\theta) &= -\frac{(1-a^2)(r-2M)r^3\sin^2\theta}{(r-M)^2}, \quad
\omega_0(r,\theta) = 0,
\end{align} 
and 
\begin{equation}
    \gamma_0(r,\theta)=-\frac{1}{2}\ln\left[ \frac{4M^2(r-M)^4}{r^6(r-2M)^2} + \frac{(r-M)^2}{r^4\sin^2\theta} \right],
\end{equation}
\end{subequations}
where the subscript $0$ denotes seed quantities.

Due to the staticity of the seed metric and the absence of charges, $\chi_0=0$, and therefore we can identify 
the seed Ernst potentials
\begin{equation}
\ernst_0=f_0, \quad \Phi_0=0.
\end{equation}
The new Ernst potential $\ernst$, the one obtained by applying the Ehlers transformation~\eqref{ehlers}, is now characterised by a nonzero imaginary part $\chi$. This is an intrinsic feature of the transformation, and it gives rise, via Eq. \eqref{chi}, to an angular velocity $\omega$, ergo, to the rotating character of the transformed background. 
Hence, 
\begin{equation}
\ernst=f+i\chi, \hspace{0.3cm} \Phi=0,
\end{equation}
where 
\begin{subequations}
\begin{align}
f(r,\theta) &\equiv \Re \ernst= \frac{f_0}{F}, \\
\chi(r,\theta) &\equiv \Im\ernst= -\jmath \frac{f_0^2}{F} \, 
\end{align}
\end{subequations}
and where we have defined 
\begin{equation}
F(r,\theta) := 1 + \jmath^2 f_0^2 \,.
\end{equation}
Knowing $\chi$, we can integrate~\eqref{chi} for the angular velocity $\omega$. However, bear in mind that what before was the flat gradient in cylindrical coordinates now acquires the nontrivial form
\begin{equation}
    \vec{\nabla} k(r,\theta) = \frac{r-M}{(1-a^2)\sqrt{r^2(r-2M)^2 + 4M^2(r-M)^2\sin^2\theta}}\left[\vec{e}_r(r-M)\partial_r k + \vec{e}_\theta\partial_\theta k\right],
\end{equation}
for any scalar function $k(r,\theta)$. Consequently, it turns out that
\begin{equation}
\omega(r,\theta) = -4\jmath (1-a^2) \frac{r^2-3Mr+3M^2}{r-M} \cos\theta \,.\label{eq:omega}
\end{equation}

The construction of our swirling geometry in the minimally coupled theory (\ref{FJNWaction}) is then completed; it only remains to move back to the conformal frame. Recalling that Ehlers transformations do not affect the metric function $\gamma$, we can substitute the previously found functions $f,\omega,\gamma\equiv\gamma_0$ into the magnetic LWP metric~\eqref{lwp-m} and apply the corresponding Bekenstein map to the solution in order to reach the final form of our Barcel\'o-Visser Swirling (BVS) wormhole, namely,
\begin{subequations}\label{swirlingWH}
\begin{align}\label{swirlingWHmetric}
d\hat{s}^2 &= \Omega \left\lbrace
F \left[
-\left(1 - \frac{M}{r}\right)^2 {dt}^2 + \frac{{dr}^2}{\left(1 - \frac{M}{r}\right)^2} + r^2 {d\theta}^2 \right]+ \frac{r^2\sin^2\theta}{F} \left( d\vartheta - \omega dt \right)^2 \right\rbrace,\\
\varphi &= \sqrt{3} \frac{M + a(r-M)}{r-(1-a)M},
\end{align}
\end{subequations}
\begin{subequations}
where $\Omega$ is given in~\eqref{eq:Omega}, $\omega$ is in~\eqref{eq:omega}, and 
\begin{align}
F(r,\theta) &= 1 + \jmath^2 (1-a^2)^2 \frac{r^6(r-2M)^2}{(r-M)^4} \sin^4\theta \,.
\end{align}
\end{subequations}
Here, and in analogy with the seed metric, the parameter $M$ represents the mass of the configuration, while on the other hand the new parameter $j$ is associated with the swirling rotation.  

\subsubsection{Geometric analysis}

The BVS spacetime describes a rotating wormhole obtained by embedding the otherwise static (uncharged) BV wormhole in a swirling spacetime. The first step in our analysis is to determine the radius of the wormhole's throat which connects the two asymptotic regions, the one located at $r=M$ and the other at radial infinity. 

The radius we are looking for is the value of the radial coordinate minimizing the surface area. Therefore, let us fix the radius as $r=r_\circ$ and compute the surface area via 
\begin{equation}
    \mathcal{A} = \Omega\,r_\circ^2 \int_0^{\pi}\sin\theta d\theta\int_0^{2\pi}d\vartheta=4\pi \,\Omega\, r_\circ^2.
\end{equation}
Given that $r>M$, the radius minimizing the area is 
\begin{equation}
    r_\circ = (1+\sqrt{a})M,\label{eq:BVS r0}
\end{equation}
with the minimum being 
\begin{equation}
    \mathcal{A}_{\text{min}} = 4\pi M^2 (1+\sqrt{a})^4.
\end{equation}
Interestingly, the radius of the throat is not affected by the rotation. This can be seen by killing the charges in \eqref{eq:BVthroat} and comparing the result to~\eqref{eq:BVS r0}.

Moreover, the wormhole is traversable. A convenient radial null observer at the equator and at a fixed arbitrary azimuthal angle, parametrized by 
\begin{equation}
    \frac{dt}{dr}=\left(1-\frac{M}{r}\right)^{-2},
\end{equation}
can traverse the throat in a finite amount of time. Indeed, the time it takes to cover the distance between $r_\circ+\epsilon$ and $r_\circ-\epsilon$ reads 
\begin{equation}
    \delta t= \int_{r_\circ-\epsilon}^{r_\circ + \epsilon}dr \frac{r^2}{(r-M)^2} = 2\left( \epsilon + \frac{\epsilon M^2}{a M^2 - \epsilon^2}+M\ln \frac{\sqrt{a} M+\epsilon}{\sqrt{a} M -\epsilon}\right),
\end{equation}
with $\sqrt{a}M>\epsilon>0$. We remark here that the 3-metric at the equator is a static metric with $\omega=0$. 

When the mass parameter $M$ vanishes, we are left with the swirling solution,
\begin{equation}
    ds^2 \underset{M=0}{=} (1+\tilde{\jmath}^2\rho^4)(-dt^2+d\rho^2+dz^2)+\frac{\rho^2}{1+\Tilde{\jmath}^2\rho^4}(d\vartheta + 4 \Tilde{\jmath}z dt)^2,\label{eq:HSback}
\end{equation}
displayed in cylindrical coordinates, where $\Tilde{\jmath}=\jmath (1-a^2)$ drives the angular velocity of the background. This background exists in the presence of a constant scalar field,
\begin{equation}
    \varphi \underset{M=0}{=} \sqrt{3}a,
\end{equation}
and thus, we may refer to this solitonic background as \emph{hairy swirling}. The pure swirling geometry discussed in Refs.~\cite{Astorino:2022aam,Gibbons:2013yq} is recovered if we further take $a=0$. Due to the hairy swirling background~\eqref{eq:HSback}, it makes sense to consider $\Tilde{\jmath}$, instead of $\jmath$, as the parameter controlling the intensity of rotation. It is also important to remark that the rotation of the BVS wormhole, due to the rotating background, is quite different from a Kerr-like type of rotation which utterly disappears when the mass vanishes. 

One can also straightforwardly verify that our full solution has the same asymptotic behavior at infinity as the hairy swirling background. Moreover, it is devoid of curvature singularities, at least those discernible through the Kretschmann scalar. Indeed, $r=M$ proves to be a coordinate singularity, for $\hat{r}=r-M$, 
\begin{equation}
    R^{\lambda\rho}{}_{\mu\nu}R^{\mu\nu}{}_{\lambda\rho} \underset{\hat{r}\to 0}{=} \frac{192\hat{r}^{12}}{a^4\Tilde{\jmath}^4 M^{24}\sin^{12}\theta} +\mathcal{O}(\hat{r}^{13}) ,\label{eq:KretschM}
\end{equation}
while $\varphi \to \sqrt{3}/a$ as $\hat{r}\to 0$. About very large $r$, we have that 
\begin{equation}
    R^{\lambda\rho}{}_{\mu\nu}R^{\mu\nu}{}_{\lambda\rho} \underset{r\to \infty}{=} \frac{192}{\Tilde{\jmath}^4 r^{12} \sin^{12}\theta }+\mathcal{O}(r^{-13}),\label{eq:KretschInf}
\end{equation}
whereas 
\begin{equation}
    \lim\limits_{r\to \infty}\varphi = \sqrt{3}a.
\end{equation}
On the other hand, evaluating the Kretschmann scalar at the poles and expanding the result about small $\hat{r}$, we find that 
\begin{equation}
    \left(R^{\lambda\rho}{}_{\mu\nu}R^{\mu\nu}{}_{\lambda\rho}\right)_{\theta=0,\pi}\underset{\hat{r}\to 0}{=}-\frac{192 \Tilde{\jmath}^2}{a^4}+\mathcal{O}(\hat{r}).
\end{equation}
If we instead expand in inverse powers of $r$, we get 
\begin{equation}
    \left(R^{\lambda\rho}{}_{\mu\nu}R^{\mu\nu}{}_{\lambda\rho}\right)_{\theta=0,\pi}\underset{{r}\to \infty}{=}-192 \Tilde{\jmath}^2+\mathcal{O}(r^{-1}).
\end{equation}
Therefore, in the two asymptotic regions, and on what would be the $z$ axis in cylindrical coordinates, spacetime seems to be of negative uniform curvature. 

Let us now visualize the wormhole by taking a constant-time slice at the equator and embedding the resulting 2-metric in three-dimensional Euclidean space. In cylindrical coordinates $\{t,\rho,z,\vartheta\}$ with
\begin{equation}
    \rho=(r-M)\sin\theta\quad \text{and}\quad z=(r-M)\cos\theta,\label{eq:CYcoords}
\end{equation}
this 2-metric reads 
\begin{equation}
    d\hat{s}^2_2 = W\left( d\rho^2 + \frac{\rho^{10}}{\left[ \rho^4 + \Tilde{\jmath}^2(\rho-M)^2(\rho+M)^6 \right]^2}d\vartheta^2\right),\label{eq:2-metric for embedding}
\end{equation}
where 
\begin{equation}
    {W}(\rho) = \frac{(\rho+M)^2(\rho+a M)^2\left[ \rho^4 + \tilde{\jmath}^2(\rho-M)^2(\rho+M)^6 \right]}{\rho^8}.
\end{equation}
The metric describing Euclidean 3-space in cylindrical coordinates $\{\mathring{\rho},\mathring{z},\vartheta\}$ is given by
\begin{equation}
    ds^2_{\text{E}} = d \mathring{z}^2 + d\mathring{\rho}^2 + \mathring{\rho}^2d\vartheta^2,\label{eq:Euclidean 3-metric}
\end{equation}
where $\mathring{z}$ gives the height of the cylinder, $\mathring{\rho}$ gives its radius, and $\vartheta$ is the azimuthal angle. 

We can always write Eq.~\eqref{eq:Euclidean 3-metric} as 
\begin{equation}
    ds^2_{\text{E}} = \left[ \left( \partial_\rho \mathring{z} \right)^2 + \left( \partial_\rho \mathring{\rho} \right)^2 \right] d\rho^2 + \mathring{\rho}^2 d\vartheta^2.
\end{equation}
Comparing it to Eq.~\eqref{eq:2-metric for embedding}, we can identify 
\begin{eqnarray}
\mathring{\rho}&=&\frac{\rho(\rho+M)(\rho+a M)}{\sqrt{\rho^4 + \Tilde{\jmath}^2(\rho-M)^2(\rho+M)^6}},\\
\partial_\rho \mathring{z}&=&\pm \sqrt{W-\left( \partial_\rho \mathring{\rho} \right)^2}=\pm\frac{\sqrt{{P}}}{\rho^4\left[ \rho^4 + \tilde{\jmath}^2(\rho-M)^2(\rho+M)^6 \right]^{3/2}},\label{eq:ODE}
\end{eqnarray}
where 
\begin{equation}
    {P} = \Tilde{\jmath}^8\rho^{36}+2M(9+a)\Tilde{\jmath}^8\rho^{35}+...+a^2M^{36}\Tilde{\jmath}^8
\end{equation}
is an utterly ugly polynomial in $\rho$ which we do not bother to write down in detail since we are going to handle things numerically after all. 

Nevertheless, it is worth mentioning that in the slow- and fast-rotation approximations we can analytically integrate Eq.~\eqref{eq:ODE}. Particularly in the simpler case of fast rotation, expanding in inverse powers of $\Tilde{\jmath}$, we get a simple differential equation, namely,
\begin{equation}
    \pm\partial_\rho \mathring{z} \underset{\Tilde{\jmath}\to \infty}{=} \frac{|\rho-M|(\rho+M)^4(\rho+aM)}{\rho^4}\Tilde{\jmath} + \mathcal{O}(\Tilde{\jmath}^{-1}).
\end{equation}
This gives us the piecewise result
\begin{equation}
    \mathring{z}=\pm\Tilde{\jmath}\begin{cases}
        -{Z}+c_1, &\text{if}\quad 0<\rho< M \\
        {Z}+c_2, &\text{if}\quad \rho \geq M,
    \end{cases}
\end{equation}
with
\begin{eqnarray}
    {Z} &=&\frac{M^4}{6\rho^3}\left[ 2 a M^2 + 3(1+3a)M\rho + 6(3+2a)\rho^2 \right] -2(1-a)M^3\ln\rho+\nonumber\\
    &&+\frac{\rho}{6}\left[ 6(2+3a)M^2+3(3+a)M\rho +2\rho^2 \right],
\end{eqnarray}
and integration constants $c_1,c_2$ fixed as 
\begin{equation}
    c_1={Z}(\sqrt{a}M)\quad \text{and}\quad c_2={Z}(\sqrt{a}M)-2{Z}(M),
\end{equation}
in order to ensure continuity and to have $\mathring{z}$ vanishing at the throat radius $\rho_\circ = \sqrt{a}M$. 

\begin{figure}
     \centering
     \begin{subfigure}[b]{0.3\textwidth}
         \centering
         \includegraphics[width=\textwidth]{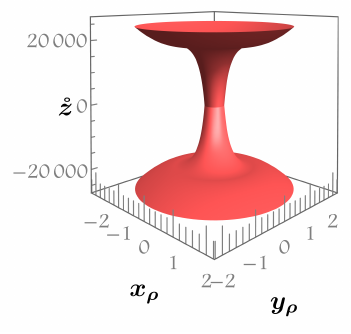}
         \caption{$a=0.0625$}
         \label{fig:a1}
     \end{subfigure}
     \hfill
     \begin{subfigure}[b]{0.3\textwidth}
         \centering
         \includegraphics[width=\textwidth]{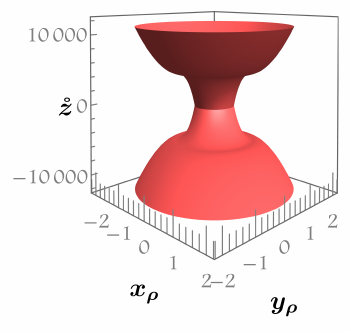}
         \caption{$a=0.25$}
         \label{fig:a2}
     \end{subfigure}
     \hfill
     \begin{subfigure}[b]{0.3\textwidth}
         \centering
         \includegraphics[width=\textwidth]{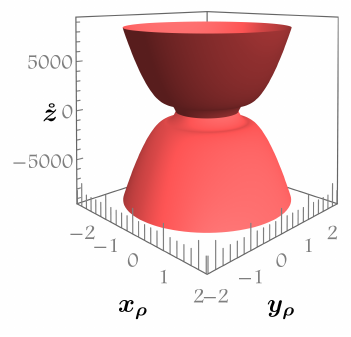}
         \caption{$a=0.5625$}
         \label{fig:a3}
     \end{subfigure}
        \caption{Embedding of a \emph{fast-rotating} BVS wormhole for different values of $a$ in three-dimensional space. We have set $M=1$ and $\Tilde{\jmath}=1000$. The variables $x_\rho$ and $y_\rho$ are given by $\rho\sin\vartheta$ and $\rho\cos\vartheta$, respectively.}
        \label{fig:BVSembedding}
\end{figure}

In Fig.~\ref{fig:BVSembedding}, we showcase visuals of a BVS wormhole for different values of $a$, considering a fast-rotation approximation. As we increase $a$, we see the characteristic increase in the radius of the wormhole's throat. Of course, an increase in the mass for fixed $a$ would produce a similar effect. Nevertheless, changing $a$ has another effect as well; as $a$ increases, the value of the embedding function $\mathring{z}$ at the inflection point $\rho=M$ (where the concavity changes) decreases. This produces the observed effect of ``two bowls approaching each other.'' The full solution can be studied numerically, if one bears in mind that the embedding equation makes sense only for $a,\tilde{\jmath},M$ such that $P>0$. For fixed $\Tilde{\jmath}$, a change in $a$ has the same effect as in the fast-rotation approximation. On the other hand, for fixed $a$, one finds that an increase in $\Tilde{\jmath}$ produces a prolongation effect which preserves the throat radius; we may loosely imagine this as ``pushing the two bowls further apart.''

Since the rotation is perhaps the most notable feature of the BVS wormhole, it is worth saying a few words about the angular velocity $\omega$. On the axis of rotation, i.e., $\rho=0$, $\omega$ assumes the form
\begin{equation}
    \omega \underset{\rho=0}{=}-\frac{4\Tilde{\jmath}}{z}\left( M^2 + z^2 - M|z|\right),
\end{equation}
meaning that $\omega\sim -4\Tilde{\jmath}z$ for large $|z|$. In other words, the angular velocity is not constant on the $z$ axis; it increases in opposite directions without upper or lower bounds.\footnote{Something similar happens with the magnetized charged black holes in Ref.~\cite{Gibbons:2013yq}.} Do also note that there is a mass-dependent single pole as both $\rho$ and $z$ go to zero, which amounts to the $r\to M$ limit in Schwarzschild coordinates. Therefore, we see that $\omega$ grows arbitrarily large in the vicinity of both conformal and radial infinity. This behavior of the angular velocity, though certainly peculiar, seems to be quite sensible in the light of the following compelling proposal presented in Ref.~\cite{Astorino:2022aam}: if we are to understand the Melvin spacetime as a system of magnetically charged Reissner-Nordström black holes displaced towards infinity, then (in an analogous fashion), we may as well understand the swirling background as a pair of counter-rotating Taub-NUT black holes located at antipodal infinities, whose combined influence generates the observed rotation. This interpretation finds support in the aforementioned behavior of the angular velocity at radial infinity, where the increase in opposite directions in the two hemispheres is consistent with the rotational directions of the individual black holes. 

\begin{figure}
     \centering
     \begin{subfigure}[b]{0.45\textwidth}
         \centering
         \includegraphics[width = \textwidth]{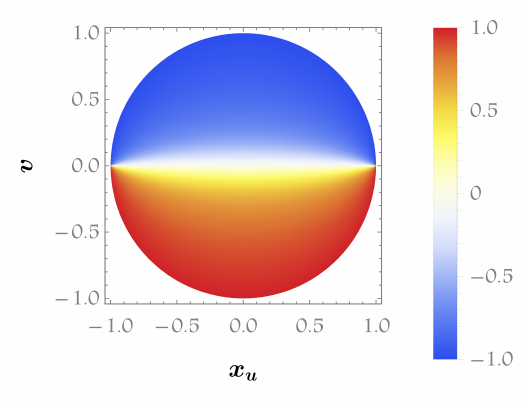}
         \caption{Hairy swirling background}
         \label{fig:DPS}
     \end{subfigure}
     \hfill
     \begin{subfigure}[b]{0.45\textwidth}
         \centering
         \includegraphics[width = \textwidth]{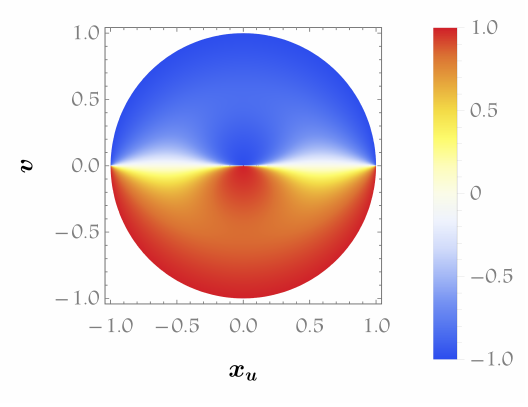}
         \caption{BVS wormhole}
         \label{fig:DPBVS}
     \end{subfigure}
    \caption{Cross section of a 3D density plot of $\tilde{\omega}:=(2/\pi)\arctan\omega$ as a function of $v$, $x_u=u\sin\vartheta$, and $y_u = u\cos\vartheta$, with $y_u=0$ and $M=\Tilde{\jmath}=1$. The angular velocity vanishes only at the equator $v=0$. The value $\tilde{\omega}=1(\tilde{\omega}=-1)$ corresponds to positive (negative) infinity for $\omega$. The three-dimensional heat map is obtained via a full revolution around the $v$ axis.}
    \label{fig:AngVel}
\end{figure}

However, this does not explain the singular behavior at the other end of the radial domain, namely, as $r\to M$. To understand the issue better, we switch to coordinates 
\begin{equation}
    u=\frac{2}{\pi}\arctan(r-M)\sin\theta\quad\text{and}\quad v=\frac{2}{\pi}\arctan(r-M)\cos\theta,\label{eq:UVcoords}
\end{equation}
for which it holds that $0\leq u < 1$ and $-1< v< 1$ with $u^2+v^2<1$. The asymptotic region $r=M$ is localized at the origin $u=0=v$, while radial infinity is at $u^2+v^2=1$, which constitutes the boundary surface. In the new coordinate system, 
\begin{equation}
    \omega(u,v) = \frac{4\Tilde{\jmath}v}{\sqrt{u^2+v^2}}\left( M - M^2 \cot \frac{\pi \sqrt{u^2+v^2}}{2} - \tan \frac{\pi \sqrt{u^2+v^2}}{2}\right).
\end{equation}
It is evident that the wormhole does not rotate at the equator $v=0$. Moreover, when both $u$ and $v$ simultaneously go to zero, we have that $\omega$ diverges as
\begin{equation}
    \omega \underset{u,v\to 0}{\sim} -\frac{8\Tilde{\jmath}vM^2}{\pi(u^2+v^2)},
\end{equation}
whereas sending $u^2+v^2\to 1$ while keeping $v$ fixed yields another divergence, namely,
\begin{equation}
    \omega \underset{u^2+v^2\to 1}{\sim} -{4\Tilde{\jmath}v}\tan \frac{\pi \sqrt{u^2+v^2}}{2}
\end{equation}
Therefore, as previously anticipated, $\omega$ becomes infinite at radial infinity due to the hairy swirling background, whereas it also becomes infinite at conformal infinity in the presence of a mass. A suggestive heat map of a convenient functional of the angular velocity is displayed in Fig.~\ref{fig:AngVel}. Bear in mind that the darker the color, the larger $|\omega|$ is. Quite interestingly then, we see that, in the spirit of the previous analogy, a configuration with two counter-rotating pairs of two corotating sources pushed toward the two asymptotic regions would exactly explain the behavior of the angular velocity in the BVS case.

Concluding, the equation describing the frame dragging effect across the entire spacetime, $\dot{\vartheta} = \omega$ (the dot stands for a derivative with respect to $x^0\equiv t$), shows that the gravitational dragging can easily surpass the value of $c=1$ (speed of light in natural units), potentially violating causality. Nevertheless, for curves with constant $t,r,\theta$, the corresponding interval reads
\begin{equation}
    d\hat{s}^2 \underset{t,r,\theta = \text{const.}}{=}\frac{\Omega}{F}r^2\sin^2\theta.
\end{equation}
Since the right-hand side is a positive expression, the above interval is spacelike, and closed timelike curves do not emerge; causality concerns are consequently resolved. Actually, a change of coordinates to a system adapted to timelike observers should show that the infinite growth is a coordinate issue after all. However, this is out of the scope of this paper. 

\begin{figure}
     \centering
     \begin{subfigure}[b]{0.3\textwidth}
         \centering
         \includegraphics[width=\textwidth]{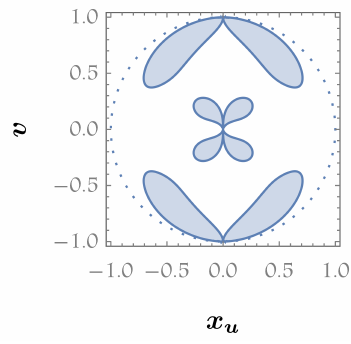}
         \caption{$\Tilde{\jmath}=0.1$}
         \label{fig:jj1}
     \end{subfigure}
     \hfill
     \begin{subfigure}[b]{0.3\textwidth}
         \centering
         \includegraphics[width=\textwidth]{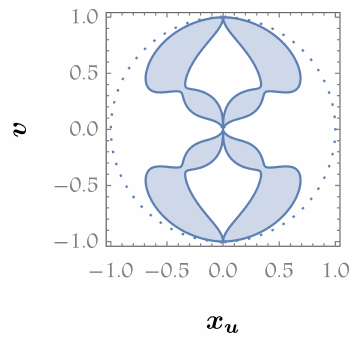}
         \caption{$\Tilde{\jmath}=0.126$}
         \label{fig:jj2}
     \end{subfigure}
     \hfill
     \begin{subfigure}[b]{0.3\textwidth}
         \centering
         \includegraphics[width=\textwidth]{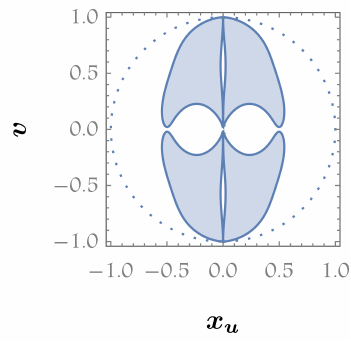}
         \caption{$\Tilde{\jmath}=1.5$}
         \label{fig:jj3}
     \end{subfigure}
        \caption{Cross section of the ergoregions of the BVS wormhole in coordinates $v$, $x_u=u\sin\vartheta$, and $y_u=u\cos\vartheta$ with $y_u=0$ and $M=1$, $a=0.5$. Dotted circle denotes the asymptotic boundary corresponding to radial infinity in Schwarzschild coordinates. The full ergoregions are obtained via a full revolution around the $v$ axis.}
        \label{fig:BVSergo}
\end{figure}

Another major feature of our rotating wormhole is the emergence of ergoregions. These are regions where the norm of the Killing vector $K=\partial_t$ is spacelike, i.e., $K^\mu K^\nu g_{\mu\nu}\equiv g_{tt}>0$. It is easy to see that, in terms of the coordinates~\eqref{eq:CYcoords}, expanding in inverse powers of $z$ while keeping $\rho$ small, we have 
\begin{equation}
    \hat{g}_{tt}\underset{z\to \infty}{=} \frac{(4\tilde{\jmath}\rho z)^2}{1+\tilde{\jmath}^2\rho^4}+\mathcal{O}(z).
\end{equation}
The leading term is clearly positive, and this signifies the presence of an ergoregion which develops toward infinite $z$.\footnote{The situation for negative $z$ is identical.} The fact that the mass has no significant effect to leading order implies that this behavior can be attributed to the underlying swirling geometry. However, numerical evidence suggests that there is also a mass-dependent ergoregion for certain values of $\tilde{\jmath}$ which develops toward conformal infinity, what would be $r=M$. Of course, since infinities are once again involved in the discussion, it is convenient to adopt the peculiar coordinates~\eqref{eq:UVcoords} to display a cross section of the ergoregions of the BVS wormhole in Fig.~\ref{fig:BVSergo}.

From the provided figure, it becomes apparent that we can identify two topologically distinct configurations. The first configuration exhibits four disjoint ergoregions, while the second configuration features only two. In the case of relatively small (but nonzero) $\Tilde{\jmath}$, the slow rotation of the hairy background leads to the emergence of two ergoregions that infinitely extend toward $r=M$. Additionally, two ergoregions appear close to the rotational axis, but not directly on it, and develop toward radial infinity (this would be $v=\pm 1$ with $u$ small in this case). Although the ergosurfaces seem to touch the axis of rotation in the above plots, this does not actually happen; they rather develop around it. As we increase the parameter $\Tilde{\jmath}$, the first configuration gradually transforms into the second configuration through what looks like a ``merging'' process. The result of this process is the emergence of two ergoregions that extend toward $v=\pm 1$, which can be interpreted as extending toward radial infinity while maintaining a relatively small value for $r\sin\theta$.

\subsection{Magnetized wormholes}

To construct a charged rotating wormhole, we proceed in a similar fashion. We embed the electrically charged BV wormhole into a Melvin universe. The final configuration rotates due to the electromagnetic interaction between the wormhole's electric charge and the magnetic background field. Again, by following the procedure of the previous section, we first perform an inverse Bekenstein transformation to land on the minimally coupled theory. The metric, scalar, and gauge fields read 
\begin{subequations}\label{CBVEF} 
\begin{align}
ds^2 &=\left[1 - \left(\frac{\sqrt{M^2-\frac{Q^2}{1-a^2}} + a(r-M)}{a\sqrt{M^2-\frac{Q^2}{1-a^2}} + r-M}\right)^2\right]  
\Omega\, d\bar{s}^2,\\ 
\phi &= \sqrt{3} \, \arctanh\biggl(\frac{\sqrt{M^2 - \frac{Q^2}{1-a^2}} + a(r-M)}{a\sqrt{M^2 - \frac{Q^2}{1-a^2}} + r-M}\biggr), \,\\ 
A &= \frac{Q}{r} \, dt ,
\end{align}
\end{subequations}
respectively, where 
\begin{equation}
    \Omega(r) = \left( \frac{a \sqrt{M^2 - \frac{ Q^2}{1 - a^2}} + r -  M}{r-M} \right)^2.\label{eq:OmegaBVM}
\end{equation}
The various functions of the seed metric are again found by comparison with the doubly Wick-rotated LWP metric~\eqref{lwp-m}. After switching coordinates to $\{t,r,\theta,\vartheta\}$ with
\begin{subequations}\label{eq:RhoZMagnetic}
\begin{eqnarray}
\rho(r,\theta) &=& \frac{(1-a^2)(r-2M)r +Q^2}{r-M} \sin\theta,\\
z(r,\theta) &=& \frac{(1-a^2)(r^2 -2 M r + 2M^2)-Q^2}{r-M}\cos\theta,
\end{eqnarray}
\end{subequations}
we obtain the identifications
\begin{subequations}
\begin{align}\label{eq:f0Magnetic}
f_0(r,\theta) &= -\frac{(1-a^2)(r-2M)r^3 +Q^2r^2}{(r-M)^2} \sin^2\theta,\quad \omega_0(r,\theta)=0,
\end{align}
and 
\begin{equation}
    \gamma_0(r,\theta)=-\frac{1}{2}\ln\left[ \frac{4(1-a^2)\left[ (1-a^2)M^2-Q^2 \right](r-M^4)}{\left[ Q^2 +(1-a^2)(r-2M)r\right]^2 r^4} + \frac{(r-M)^2}{r^4 \sin^2\theta} \right].
\end{equation}
\end{subequations}
In terms of the new coordinates, the gradient reads 
\begin{equation}
    \vec{\nabla}k(r,\theta) = \frac{(r-M)\left(\vec{e}_r (r-M)\partial_r k + \vec{e}_\theta \partial_\theta k \right)}{\sqrt{\left[ Q^2+(1-a^2)(r-2M) r \right]^2 +4(1-a^2)\left[ (1-a^2)M^2 - Q^2\right](r-M)^2\sin^2\theta}},
\end{equation}
for any scalar function $k(r,\theta)$. We can integrate Eq.~\eqref{At} for the seed twisted potential $\tilde{A}_{t0}$, which obviously cannot depend on $r$. Indeed, 
\begin{equation}
    \tilde{A}_{t0}(\theta) = Q\cos\theta,
\end{equation}
and we can identify the seed Ernst potentials 
\begin{subequations}
\begin{align}
\ernst_0(r,\theta) &= f_0 - Q^2 \cos^2\theta \,, \\
\Phi_0(r,\theta) &= iQ \cos\theta \,.
\end{align}
\end{subequations}
Notice that the imaginary part of $\ernst_0$ is zero due to the static nature of the seed metric and the absence of a magnetic component in the definition of the seed gauge field which results in a purely real $\Phi_0^\ast \vec{\nabla}\Phi_0$.

The new Ernst potentials are given by~\eqref{harrison}, where we have considered $\alpha=B/2$ as a \emph{real} parameter. Defining 
\begin{equation}
\Lambda(r,\theta) := 1 - \frac{B^2}{4} \mathcal{E}_0-iBQ \cos\theta \,,
\end{equation}
we express the transformed potentials as
\begin{subequations}
\begin{align}
\ernst(r,\theta) &= \frac{\ernst_0}{\Lambda} \,, \\
\Phi(r,\theta) &= \frac{\Phi_0 + \frac{B}{2}\ernst_0}{\Lambda} \,.
\end{align}
\end{subequations}
Therefore, we can identify 
\begin{subequations}\label{eq:NewFunctionsMagnetic}
    \begin{eqnarray}
        A_\vartheta(r, \theta) &=&  \frac{B}{8|\Lambda|^2} \left[ \ernst_0\left( 4-\ernst_0B^2\right) - 8Q^2\cos^2\theta\right],\\
        \tilde{A}_t(r,\theta) &=& \frac{Q(4+B^2\ernst_0)\cos\theta}{4|\Lambda|^2},\\
        \chi(r,\theta) &=& \frac{BQ\ernst_0 \cos\theta}{|\Lambda|^2},\\
        f(r,\theta) &=& \frac{f_0}{|\Lambda|^2}.
    \end{eqnarray}
\end{subequations}
Having these at hand, we can obtain the functions $\omega$ and $A_t$ by integrating~\eqref{chi} and ~\eqref{At}, finding
\begin{subequations}
\begin{align}
 \omega(r, \theta) &= \frac{BQ}{2 r (r - M)} \left\lbrace 4M - r\left[ 4+B^2\left( Q^2 - (1-a^2)r^2\right)\right]\right.+\nonumber \\  
   & + \left.B^2\left[ (1-a^2)(r^2+3M^2)r- M\left( Q^2 +3(1-a^2)r^2\right) \right]\cos^2\theta\right\rbrace,\label{eq:OmegaMagnetized}\\
   A_t(r,\theta)&=-\omega\left( A_\vartheta + \frac{3}{2B}\right)-\frac{2Q}{r}.
\end{align}
\end{subequations}

Once again, the final solution is found via a Bekenstein transformation that brings us back to the conformal frame, yielding the magnetized BV (MBV) wormhole
\begin{subequations}\label{MelvinWH}
\begin{align}
d\hat{s}^2 &= \Omega \left\lbrace
|\Lambda|^2 \left[
-\left(1 - \frac{M}{r}\right)^2 {dt}^2 + \frac{{dr}^2}{\left(1 - \frac{M}{r}\right)^2} + r^2 {d\theta}^2 \right] + \frac{r^2\sin^2\theta}{|\Lambda|^2} \left( d\vartheta - \omega dt \right)^2 \right\rbrace,\\
\varphi &= \sqrt{3}  \frac{\sqrt{M^2 - \frac{Q^2}{1-a^2}} + a(r-M)}{a \sqrt{M^2 - \frac{Q^2}{1-a^2}} + r-M} \,,\\
A &= A_t dt + A_\vartheta d\vartheta \,, 
\end{align}
\end{subequations}
where all functions involved have been previously defined in this section.

\subsubsection{Geometric analysis}
The MBV spacetime describes a wormhole rotating due to electromagnetic interactions. It is obtained by embedding the static electrically charged BV wormhole into a Melvin universe. One can straightforwardly verify that the throat of this wormhole is exactly of radius~\eqref{eq:BVthroat} and that the wormhole is traversable with $\delta t$ given by~\eqref{eq:travBVstatic}.

Do further observe that 
\begin{equation}
    {|Q|}\leq \sqrt{1-a^2}M,
\end{equation}
in order for the conformal factor to be real, and thus, the charge of the solution cannot be switched off in the presence of a mass.\footnote{Actually, the equality is not of interest because it pushes the throat radius to conformal infinity $r=M$.} Indeed, the true background should be for $M=0=Q$, and not surprisingly, we see that it is represented by the Melvin metric \cite{Ernst:1976mzr} which in standard cylindrical coordinates reads
\begin{equation}
    ds_{\mathrm{Melvin}}^2 = \Lambda_0{}^2 \left(-dt^2 + dz^2 +d\rho^2\right) + \frac{\rho^2 d\vartheta^2}{\Lambda_0{}^2},
\end{equation}
with 
\begin{equation}
    \sqrt{|\Lambda|^2}\underset{M=0=Q}{\equiv}\Lambda_0(\rho)=1+\frac{\Tilde{B}^2 \rho^2}{4},\qquad \Tilde{B} = \sqrt{1-a^2}B. 
\end{equation}
Note, however, that this soliton is, once again, hairy, for the scalar acquires the constant profile $\varphi=\sqrt{3}a$. In contrast to the BVS, here the background is static, not just stationary.

\begin{figure}
     \centering
     \begin{subfigure}[b]{0.45\textwidth}
         \centering
         \includegraphics[width = \textwidth]{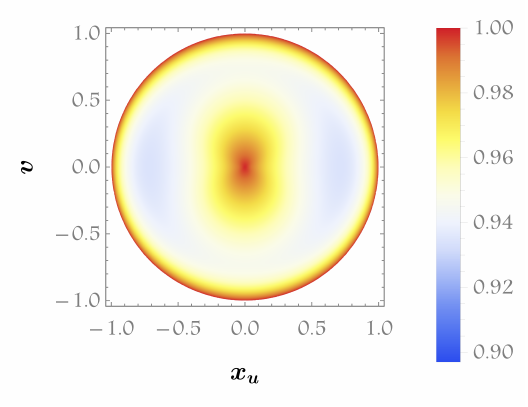}
         \caption{$M=3$}
         \label{fig:MAGhm1}
     \end{subfigure}
     \hfill
     \begin{subfigure}[b]{0.45\textwidth}
         \centering
         \includegraphics[width = \textwidth]{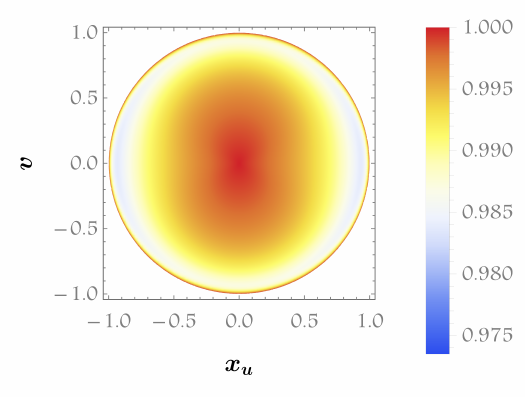}
         \caption{$M=10$}
         \label{fig:MAGhm2}
     \end{subfigure}
    \caption{$y_u=0$ cross section of a 3D density plot of $\tilde{\omega}:=(2/\pi)\arctan\omega$ as a function of $v$, $x_u=u\sin\vartheta$, and $y_u = u\cos\vartheta$, with $\Tilde{Q}=2$ and $\Tilde{B}=1$. The value $\tilde{\omega}=1$ corresponds to positive infinity for $\omega$. The three-dimensional heat map is obtained via a full revolution around the $v$ axis.}
    \label{fig:AngVelMag}
\end{figure}

Indeed, the MBV solution rotates with angular velocity which depends on $BQ$; i.e., the rotation is due the interaction of the electric charge of the charged BV solution with the background's constant magnetic field. Therefore, the wormhole cannot rotate in the absence of $Q$. Do further observe that, in contrast to the swirling case, here the wormhole rotates also at the equator $\theta=\pi/2$ with velocity 
\begin{equation}
    \omega\underset{\theta=\pi/2}{=}-\frac{BQ}{2r(r-M)}\left\lbrace r\left[ 4+B^2\left( Q^2 -(1-a^2)r^2\right) \right]-4M \right\rbrace.
\end{equation}
At this stage, it is convenient to use the cylindrical coordinates~\eqref{eq:CYcoords}. In these coordinates, we observe the following behavior,
\begin{equation}
    \omega\underset{z,\rho\to 0}{\sim} \frac{QB^3\left[ (1-a^2)M^2-Q^2 \right](2z^2+\rho^2)}{2(z^2 +\rho^2)^{3/2}},\quad \omega\underset{z,\rho\to \infty}{\sim}\frac{(1-a^2)QB^3(2z^2 + \rho^2)}{2\sqrt{z^2+\rho^2}},
\end{equation}
meaning that $\omega$ diverges near both asymptotic regions regardless of the direction from which it approaches them. To visually provide this information, it is useful to perform a final coordinate transformation to reach the system $\{t,u,v,\vartheta\}$ with $u,v$ given in~\eqref{eq:UVcoords}, and show a heat map in Fig.~\ref{fig:AngVelMag}.
We thus verify that, once again, the rotation becomes more intense close to the asymptotic regions. An increase in $\Tilde{Q}\equiv Q/\sqrt{1-a^2}=2$ and/or $\Tilde{B}\equiv B\sqrt{1-a^2}=1$ causes an overall increase of the rotational velocity. Bear in mind that the reality of the solution requires $M>\Tilde{Q}$. In contrast to the corresponding behavior in the BVS case, here the sign of $\omega$ is everywhere the same. Finally, one can easily show that (these rather spurious) infinities pose no threat to causality as there are no closed timelike curves after all. 

\begin{figure}
     \centering
     \begin{subfigure}[b]{0.3\textwidth}
         \centering
         \includegraphics[width=\textwidth]{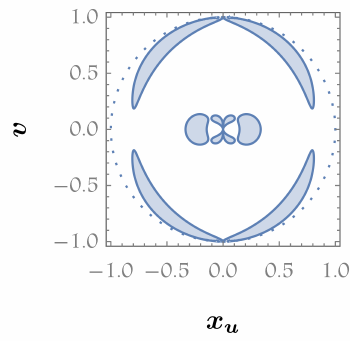}
         \caption{$\Tilde{Q}=3.95,\,\Tilde{B}=0.3$}
         \label{fig:q11}
     \end{subfigure}
     \hfill
     \begin{subfigure}[b]{0.3\textwidth}
         \centering
         \includegraphics[width=\textwidth]{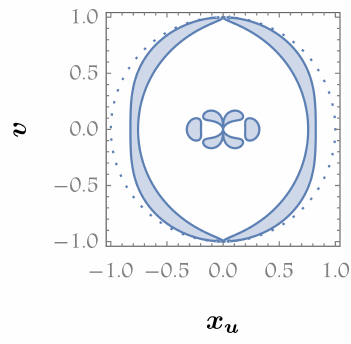}
         \caption{$\Tilde{Q}=3.93,\,\Tilde{B}=0.3$}
         \label{fig:q12}
     \end{subfigure}
     \hfill
     \begin{subfigure}[b]{0.3\textwidth}
         \centering
         \includegraphics[width=\textwidth]{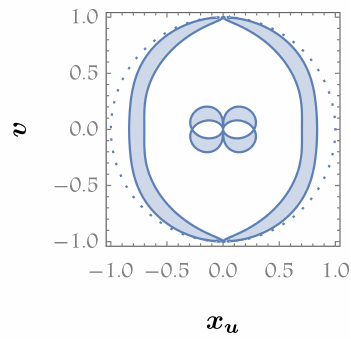}
         \caption{$\Tilde{Q}=3.9125,\,\Tilde{B}=0.3$}
         \label{fig:q13}
     \end{subfigure}\\
     \begin{subfigure}[b]{0.3\textwidth}
         \centering
         \includegraphics[width=\textwidth]{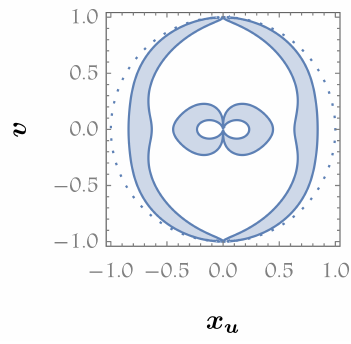}
         \caption{$\Tilde{Q}=3.9,\,\Tilde{B}=0.3$}
         \label{fig:q21}
     \end{subfigure}
     \hfill
     \begin{subfigure}[b]{0.3\textwidth}
         \centering
         \includegraphics[width=\textwidth]{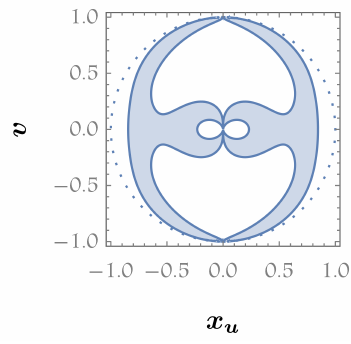}
         \caption{$\Tilde{Q}=3.89,\,\Tilde{B}=0.3$}
         \label{fig:q22}
     \end{subfigure}
     \hfill
     \begin{subfigure}[b]{0.3\textwidth}
         \centering
         \includegraphics[width=\textwidth]{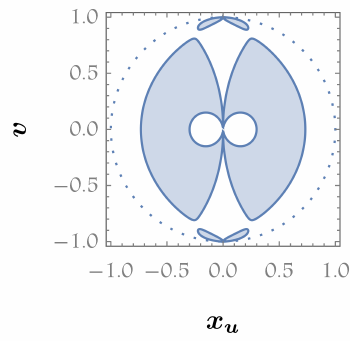}
         \caption{$\Tilde{Q}=3.89,\,\Tilde{B}=5.5$}
         \label{fig:q23}
     \end{subfigure}
        \caption{$u\cos\vartheta\equiv y_u=0$ cross section of the MBV ergoregions for different values of $\Tilde{Q},\Tilde{B}$ with $M=4$ and $a=0.5$. The full picture is given by revolution around the $v$ axis which the ergoregions never intersect.}
        \label{fig:MBVergo}
\end{figure}

Moving on, a quick inspection of $\hat{g}_{tt}$ reveals the existence of infinitely extended ergoregions close to the axis of rotation. In coordinates~\eqref{eq:CYcoords}, expanding $\hat{g}_{tt}$ in negative powers of $z$ while holding $\rho$ fixed, we find that 
\begin{equation}
    \hat{g}_{tt}\underset{z\to \infty}{\sim} \frac{\Tilde{B}^6 \Tilde{Q}^2\rho^2}{\Tilde{B}^2\Tilde{Q}^2+\frac{1}{16}(4+\Tilde{B}^2\Tilde{Q}^2 + \Tilde{B}^2\rho^2)^2},
\end{equation}
where the expression on the right can acquire arbitrarily large positive values. Nevertheless, determining the ergoregion(s) analytically is practically impossible due to the complexity of the involved functions, and thus, we resort to numerical methods to unravel the following fascinating observations using the visual aid of Fig.~\ref{fig:MBVergo}.  

In Fig.~\ref{fig:MBVergo}(\subref{fig:q11}) we find five disconnected ergoregions with two of them infinitely extending toward what would be radial infinity near the poles, two which asymptotically approach conformal infinity near the rotation axis from opposite directions and, most importantly, one ``donut-shaped'' ergoregion of \emph{finite} volume. In Fig.~\ref{fig:MBVergo}(\subref{fig:q12}), the (let us call them) outer ergoregions merge into one, while the inner topology stays the same although the finite-volume ergoregion becomes smaller. This gives us four disconnected ergoregions. Then, in Fig.~\ref{fig:MBVergo}(\subref{fig:q13}), it is the outer picture which remains the same, with the inner topology undergoing change. In particular, the finitely sized ergoregion \emph{vanishes completely}, rather than merging with the two regions of positive $\hat{g}_{tt}$ which asymptotically develop toward conformal infinity (what would be the origin in the plots). This reduces the number of ergoregions to $3$. Next, in Fig.~\ref{fig:MBVergo}(\subref{fig:q21}) the two inner ergoregions that remain, further merge into one, leaving us with two disconnected regions, which then also merge into one in Fig.~\ref{fig:MBVergo}(\subref{fig:q22}). Finally, in Fig.~\ref{fig:MBVergo}(\subref{fig:q23}), we have yet another configuration with three disconnected ergoregions, which, however, is topologically different from the one in Fig.~\ref{fig:MBVergo}(\subref{fig:q13}) in the sense that the former has two outer ergoregions and one inner, whereas the latter exhibits one outer ergoregion and two inner.

\section{Further remarks}

In this study, we have developed a novel and computationally efficient methodology for constructing exact rotating wormhole configurations. Our approach is based on the insightful description of axially symmetric and stationary spacetimes provided by Ernst, which reveals a set of otherwise hidden Lie point symmetries in Einstein-Maxwell theory. These symmetries can be wisely utilized as a solution generating technique. Specifically, we have employed magnetic Ehlers and Harrison transformations to introduce rotation into the static Barcel\'o-Visser wormhole, a static wormhole geometry structure within the Einstein-conformal-scalar system. It has been demonstrated that Ehlers symmetries permit the inclusion of a minimally coupled scalar field \cite{Astorino:2013xc} , thus expanding the spectrum of solutions within the Einstein-conformal-scalar framework through the combination of these symmetries with Bekenstein transformations \cite{Bekenstein:1974sf}. Initially, we embarked on constructing a rotating wormhole devoid of charge by incorporating the static Barcel\'o-Visser geometry into a swirling/rotating background. Recent investigations have shed light on the magnetic Ehlers transformations as an efficient technique for embedding a given seed spacetime within a rotating background \cite{Astorino:2022aam}. It is worth noting that the rotation, although bearing resemblances to standard rotating solutions in GR, does not adhere to the Kerr type, meaning it does not arise from a rotating source but rather from a rotating background that imparts its motion onto the source.
On the other hand, we extended the Barcel\'o-Visser wormhole to include rotation by placing its electrically charged counterpart \cite{Barrientos:2022avi} within an external magnetic field. This was achieved by utilizing the well-studied magnetic Harrison transformation. The rotational nature of the resulting solution arises from the interaction between the wormhole's electric charge and the external magnetic field present in the background. For both solutions, we explored various aspects including the location of the wormhole throats, traversability properties, Euclidean embeddings, asymptotic behavior, as well as the features and geometric structures of the corresponding backgrounds and ergospheres. This completes a concise geometrical analysis of our backreactions. 
Our solutions open up numerous avenues to delve into the multifaceted realm of wormhole physics. As previously emphasized, a significant challenge lies in studying the stability of these configurations. Recently, it has been demonstrated that for the class of four-dimensional wormholes within the Ellis-Bronnikov family \cite{Ellis:1973yv,Bronnikov:1973fh}, the unstable mode present in the static geometry becomes progressively more stable under increasing rotation within a perturbative regime \cite{Azad:2023iju}. At the second order in rotation, the emergence of a second mode has been observed, which tends to intertwine with the original unstable mode. It has been conjectured that rapidly rotating wormholes could attain stability. Therefore, our geometries provide exact, fully rotating wormhole spacetimes that can be utilized to test these stability analyses.
Furthermore, the pursuit of astrophysical detection of wormhole solutions has already commenced \cite{Abe:2010ap,Toki:2011zu,Takahashi:2013jqa}. Consequently, it becomes imperative to conduct a meticulous analysis of the behavior of null geodesics in our swirling and Melvin wormhole geometries to facilitate the study of their shadows \cite{Gyulchev:2018fmd,Shaikh:2018kfv,Perlick:2021aok}. This phenomenon transcends black holes and appears to be inherent in other compact objects as well. It is also natural to inquire about the gravitational lensing properties exhibited by these rotating wormholes, including the distribution of photon spheres and Einstein rings in relation to the shadows \cite{Cramer:1994qj,Safonova:2001vz,Perlick:2003vg,Nandi:2006ds,Kuhfittig:2013hva,Tsukamoto:2016zdu,Shaikh:2018oul,Jusufi:2017mav,Huang:2023yqd}. Additionally, exploring the echoes of wormholes \cite{Mark:2017dnq,Bueno:2017hyj,Liu:2020qia,Vlachos:2021weq,Yang:2022ryf} and energy extraction processes \cite{Patel:2022jbk,Ye:2023xyv}  represents among other possibilities, further avenues to test these geometries.

\section{Acknowledgments} 

We would like to thank Marco Astorino, Jutta Kunz, and Julio Oliva for their comments and suggestions. A.C. would like to acknowledge the Physics Department of University of Concepci\'on and the Institute of Mathematics of the Czech Academy of Sciences for their kind hospitality during the development of this project. The work of A.C. is funded by FONDECYT Regular Grant No. 1210500. The work of K.M. is funded by Beca Nacional de Doctorado ANID Grant No. 21231943.
K. P. acknowledges financial support
provided by the European Regional Development Fund (ERDF) through the Center of Excellence TK133 “The
Dark Side of the Universe” and PRG356 “Gauge gravity: unification, extensions and phenomenology.” K.P. also
acknowledges participation in the COST Association Action CA18108 “Quantum Gravity Phenomenology in the
Multimessenger Approach (QG-MM).”

\end{document}